# Assessment of hydrophobicity scales for protein stability and folding using energy and RMSD criteria

Boris Haimov[1] and Simcha Srebnik[*,1,2]

[1]Russell Berrie Nanotechnology Institute, Technion – Israel Institute of Technology, Haifa, 32000, Israel.

[2]Department of Chemical Engineering, Technion – Israel Institute of Technology, Haifa, 32000, Israel.

[*]Correspondence to Simcha Srebnik: simchas@technion.ac.il

**ABSTRACT**

*De novo* prediction of protein folding is an open scientific challenge. Many folding models and force fields have been developed, yet all face difficulties converging to native conformations. Hydrophobicity scales (HSs) play a crucial role in such simulations as they define the energetic interactions between protein residues, thus determining the energetically favorable conformation. While many HSs have been developed over the years using various methods, it is surprising that the scales show very weak consensus in their assignment of hydrophobicity indexes to the various residues. In this work, several HSs are systematically assessed via atomistic Monte Carlo simulation of folding of small proteins, by converting the HSs of interest into residue-residue contact energy matrices. HSs that poorly preserve native structures of proteins were tuned by applying a linear transformation. Subsequently, folding simulations were used to examine the ability of the HSs to correctly fold the proteins from a random initial conformation. Root mean square deviation (RMSD) and energy of the proteins during folding were sampled and used to define an ER-score, as the correlation between the 2-dimensional energy-RMSD (ER) histogram with 50% lowest energy conformations and the ER histogram with 50% lowest RMSD conformations. Thus, we were able to compare the ability of the different HSs to predict *de novo* protein folding quantitatively.

INTRODUCTION

Computational protein folding prediction is an extensively studied research field [1,2] that can be divided into two main branches: template based modeling (TBM), and template free modeling (TFM) that is also known as *de novo* modeling. In TBM [3] a protein is folded according to *a priori* knowledge of similar polypeptide fragments with already known folded structures usually taken from the worldwide protein data bank (PDB) [4], while no *a priori* folding knowledge is used in TFM [5]. Recent analysis by Brylinski [6] suggests that even with the continual rapid growth of PDB structures, TBM is unlikely to be sufficient for solving the folding prediction problem, and that TFM techniques need to be further developed to predict folding of unknown structures. The TFM branch includes approaches that represent systems down to atomistic detail with explicit water molecules, as well as models with reduced representation. Systems that are represented in full atomistic detail use fundamental force field equations for the simulation. Such approaches are slow and require expensive hardware. On the other hand, systems with reduced representation of proteins, implicit water molecules, and that make use of hydrophobicity scales (HSs), are faster and affordable. Since the hydrophobicity of the amino acid (AA) residues along the polypeptide play an important role in the determination of protein structure and its function, it is essential to be able to quantitatively examine and assess the underlying HS.

Hydrophobicity is a measure of how much some given object (molecule, substance, phase, structure, surface, etc.) "fears" water and generally describes the amount of energy (or some proportional value) that is required to transfer the object from non-aqueous medium into an aqueous one [7]. The object is considered to be hydrophobic if the amount of energy is greater than zero and hydrophilic otherwise. Until the 20$^{th}$ century, hydrophobicity was observed macroscopically in the form of contact angle [8–10]. After the genesis of structural biology in the middle of the 20[th] century [11], hydrophobicity scales [12–15] have been intensively used to describe the relative hydrophobicity of molecules and particularly the relative hydrophobicity of the 20 naturally occurring AAs that serve as the building blocks of proteins. The applications of HSs are numerous and include profiling protein structures [16], drug design [17], classification of protein structure [18], protein and peptide separations [19], prediction of hydrophobic cores (HCs) [20], and protein simulation [1].

To date, over 100 HSs have been developed [21,22] using a wide variety of knowledge-based [23], experimental [24], and other [25–27] methods. In most cases, HSs are described as vectors of energy (or some other proportional) values for the 20 AAs, each representing the relative hydrophobicity of the corresponding AA. Mixtures of different AA species, as normally observed in polypeptides, require a more accurate approach, where all contacts are taken into account. Thus, the HS becomes a 20x20 matrix, usually named as residue-residue (RR) contact matrix [28,29]. Surprisingly, some scales vary significantly from others and the task of picking the correct HS becomes non-trivial. Much effort has been invested in understanding the differences between the many existing HSs [21,22]. Since there are no standard units for hydrophobicity, the scales are compared by the relative order of AA hydrophobicity, and when the magnitudes of hydrophobicities are of importance a common practice is to introduce a linear transformation to the compared HS images [21].

A properly determined HS plays a crucial role in modeling of proteins using reduced representation. Evidently, the wide choice of existing HSs and their variance suggests that a systematic approach of assessing and evaluating HSs does not exist. The purpose of this study is to provide analytical means for the systematic assessment of HSs via atomistic simulation of proteins, and to analyze the ability of *de novo* protein folding prediction of several known HSs.

**METHODS**

Atomistic simulations with implicit solvent were used to assess the hydrophobicity scales (HSs) by testing the ability of the HS to preserve the folded configuration of test proteins, and via folding simulation from a random non-folded state. The force field (FF) used in the simulations is defined according to the given HS. Four hydrophobicity scales were chosen: two matrix scales (M0 [29] and M1 [28]) that are already given as residue-residue (RR) contact energy matrices, and two vector scales (V0 [30] and V1 [31]) that are converted to the form of RR contact energy matrices as described below. The HSs were picked according to a chronological criterion: two of relatively earlier works (before year 2000), and two of relatively later works (after year 2000).

Simulations are carried out using a modified Metropolis Monte Carlo (MMC) algorithm[32,33] under the canonical ensemble. Standard MMC algorithm assumes a uniform step size for all the degrees of freedom (DOFs) within the system, which is correct only for isotropic DOFs. However, the different secondary structures (e.g., helix or random coil) may have inherently different dynamics. The modified MMC algorithm used in this study assumes independent and variable step size for every DOF within the system. The step size for each DOF is continuously modified to achieve approximately 50% acceptance ratio. Simulations and analysis were carried out using in-house software that was developed under C++ and Matlab[TM]. VMD software[34] was used for the visualization of molecular structures.

POLYPEPTIDE REPRESENTATION. The polypeptide backbone, which consists of a repeating linear sequence of nitrogen (N), α-carbon (Cα), carbonyl carbon (C), and carbonyl oxygen (O), can be represented with bend, roll, and displacement (BRAD) parameters (depicted in Figure 1. The BRAD parameters were sampled directly from the PDB and averaged over all AA sequential pairs. The angles $\varphi$ and $\psi$ are known as the Ramachandran [35,36] dihedrals and define the effective conformation of the polypeptide backbone. The 3$^{rd}$ roll angle $\omega$ is essentially constant at $\omega = 180°$. Coordinates of the backbone carbonyl oxygens were obtained assuming that C$\alpha$, C, O, N backbone atoms are found on the same plane with equal bend angles $\angle$C$\alpha$-C-O = $\angle$N-C-O. Explicit hydrogen atoms were not included in the model.

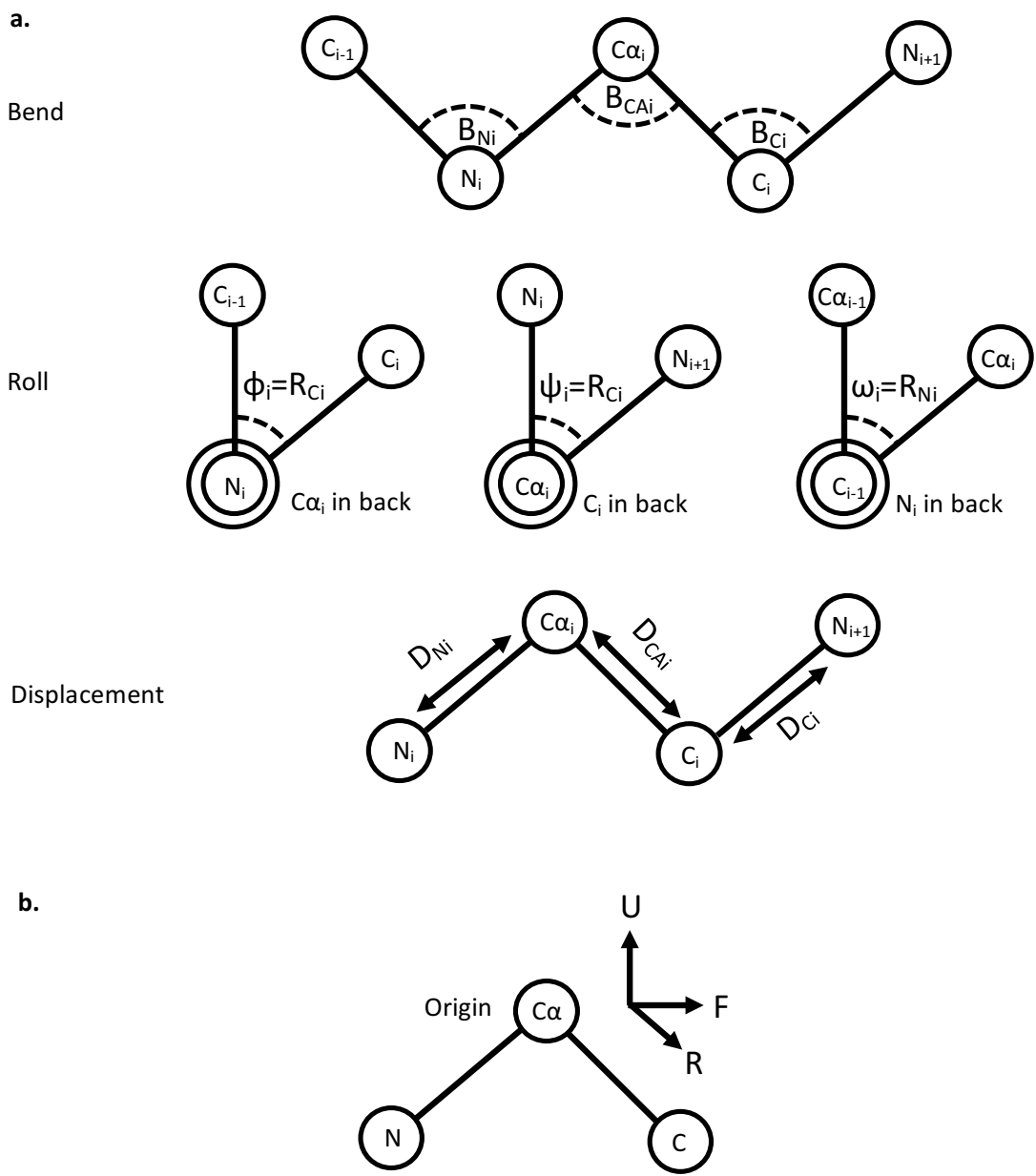

**Figure 1:** (a) Illustration of the bend ($B_N, B_{CA}, B_C$) roll ($R_{CA}, R_C, R_N$) and displacement ($D_N, D_{CA}, D_C$) (BRAD) degrees of freedom for the representation of a polypeptide backbone conformation. N stands for nitrogen atoms, Cα – alpha carbon atoms, C – carbonyl carbon atoms. Index *i*-1 and *i* refer to sequential AAs. (φ,ψ) are known as Ramachandran dihedrals and ω is a constant dihedral equal to 180˚. BRAD representation allows performing a single folding operation on polypeptide backbones with a complexity of $O<1>$, thus allowing simulation with better performance than on explicit Cartesian space of every backbone atom in the system. (b) Local up, right, and forward (URF) coordinates for the representation of residue atoms, the position of α-carbon defines the origin of the local coordinates system. Representation of residue atoms with URF coordinates allow performing manipulation on backbone conformation without the need of updating the local coordinates of every residue, thus saving CPU time.

BRAD representation, which is conceptually similar to Flory's convention [37], was chosen since it allows performing a single folding operation on polypeptide backbones (i.e., large segmental movements) with a complexity of $O<1>$, thus allowing simulation with better performance than in Cartesian space. For optimal performance, side chains are represented by local Cartesian (U,R,F) coordinates of every non-hydrogen atom with the origin defined at the Cα position, the up direction defined as U = norm ( Cα–N + Cα–C ), the right direction defined as R = norm ( U x N–Cα), and the forward direction defined as F = norm ( U x R ). The resulting URF coordinates are orthonormal and satisfy U·R=U·F=R·F=0, |U|=|R|=|F|=1, U x R = F, R x F = U, F x U = R. The main advantage of representing residue atoms with local URF coordinates is that it allows performing manipulation on the backbone conformation without the need of updating the local coordinates of every atom of the residue, thus saving CPU time. BRAD representation together with URF coordinates is a powerful combination for *in silico* handling of polypeptide conformation down-to atomic resolution.

FORCE FIELD (FF). Energetic contributions to the conformational energy of the protein include steric repulsion, hydrogen bonds (HBs), and the interaction energies between residues and tails:

$$E = E_{ST} + E_{HB} + E_{RES} + E_{TERM} \qquad (1)$$

where $E_{ST}$ is the total steric energy contributed by LJ repulsive terms, $E_{HB}$ is the total HB interaction energy, $E_{RES}$ is the total interaction energy, and $E_{TERM}$ is the contribution of termini-residue and N-terminus-C-terminus interaction energies.

To avoid sterically forbidden states we used OPLS [38,39] parameters with geometric mixing rules to calculate the repulsive energy between all non-bonded atom pairs, and for 1-3 and longer bonded neighbor interactions. A modified version of the shifted LJ-potential was used for the calculation of steric energies:

$$E_{ST}(r) = \begin{cases} \varepsilon + 4\varepsilon\left[\left(\frac{0.85\sigma}{r}\right)^{12} - \left(\frac{0.85\sigma}{r}\right)^6\right], & r < 2^{1/6} \\ 0, & r \geq 2^{1/6} \end{cases} \quad (2)$$

The factor 0.85 was applied heuristically to compensate for the increased repulsion caused by shifting the potential by ε. The applied modification ensures only repulsive contribution to the energy ($E_{LJ}(r) \geq 0$). OPLS parameters that were used for calculation of the steric energy are given in Table 1.

**Table 1: OPLS parameters for backbone and residue atoms**

|  | Backbone atoms | | | | Residue atoms | | | |
| --- | --- | --- | --- | --- | --- | --- | --- | --- |
|  | N | Cα | C | O | C | N | O | S |
| σ[Å] | 3.25 | 3.5 | 3.75 | 2.96 | 3.5 | 3.25 | 2.96 | 3.55 |
| ε[kcal/mol] | 0.17 | 0.066 | 0.105 | 0.21 | 0.066 | 0.17 | 0.21 | 0.25 |

Hydrogen bond interaction energy was calculated via:

$$E_{HB} = -4 \cdot S \text{ [kcal/mol]} \quad (3)$$

where $0 \leq S \leq 1$ is the alignment score between every amide hydrogen and carbonyl oxygen along the polypeptide backbone, introduced elsewhere [40]. For the calculation of residue-residue (RR) interaction energy, we use a 20X20 contact energy cost matrix (CECM) which defines the interaction energy of the 20 AAs. In case CECM is not given directly as a matrix, we use HS as the diagonal of the contact matrix and estimate all other contact energies as an average of the two contributing AA contact energies CECM(r,c)=HS(r)/2+HS(c)/2, where *r* and *c* correspond to the row and column AAs in the matrix. The calculation of contact energy between two residues is obtained from:

$$E_{\text{RES}} = \text{CECM}(\text{RES1}, \text{RES2}) \cdot \text{CS}(d) \qquad (4)$$

CS(*d*) is the contact score between the residues and is calculated as:

$$\text{CS}(d) = \begin{cases} 1, & d < 4 \\ 3 - 0.5d, & 4 \leq d < 6 \\ 0, & 6 \leq d \end{cases} \qquad (5)$$

where *d* is the distance between the centers of the nearest atoms of the given residues. The values of CS(*d*) range from 0 to 1. N-terminus and C-terminus are treated as lysine and aspartate residues, respectively, due to their similarity, such that:

$$E_{\text{TERM}} = \text{CECM}(\text{LYS}, \text{ASP}) \cdot \text{CS}(d_{TERM}) \qquad (6)$$

where $d_{\text{TERM}}$ is the distance between nearest atom centers of the termini residues.

EQUILIBRATION AND FOLDING SIMULATIONS. Each equilibration simulation consists of 100K MMC moves with initial conditions sampled from the PDB. Ten independent trajectories were carried out for each equilibration simulation. Folding simulations were carried out using 3 recipes: Recipe 0 begins with random conformation and proceeds with standard MMC moves. Recipe 1 begins with random conformation and proceeds with (*φ,ψ*) pairs sampled from shaped noise generator of the specific AA transition angles [40]. Recipe 2 begins with a conformation that corresponds to the most probable dihedral angles for the AA sequence and proceeds with standard MMC moves. Folding simulations start with 20K moves of the side chains only, followed by 250K moves of all atoms as specified by the recipe, for 100 independent trajectories.

**RESULTS AND DISCUSSION**

Four hydrophobicity scales (HSs) were examined for their ability to preserve the folded conformation of several small proteins, as well as their ability to capture the native conformation from an initial random conformation. The prospect of these HSs to predict the correct fold was then assessed using RMSD and energy plots and a newly defined scoring method. Table 2 presents the order of amino acids (AAs) from the least hydrophobic (top) to the most hydrophobic (bottom) according to the four chosen HSs. For the matrix scales (M0 and M1) the order was extracted from the diagonals of the residue-residue (RR) contact energy matrices. AAs with hydrophobicity values closest to zero are shown in bolded and enlarged font, such that all hydrophobicity values above bolded AA are positive, and all values below are negative. Acidic and basic AAs are marked in red and blue, accordingly, with the exception of HIS. The reason for the exclusion of HIS from the basic AA list is because its pKa is only slightly below the pH of water, so that it can be categorized as uncharged. The energies of the least hydrophobic (Max energy) and the most hydrophobic (Min energy) AAs are shown in Table 2 as well as the energy range of every HS. The differences are probably due to the different experimental approaches used for the determination of the HSs [22]. However, the evident similarity between the energy ranges of the HSs shows an interesting consensus.

**Table 2: Hydrophobicity Scales**

| | Hydrophobicity Scale | | | |
|---|---|---|---|---|
| | M0 [29] | M1 [28] | V0 [30] | V1 [31] |
| Least hydrophobic | LYS[B] | **LYS[B]** | GLU[A] | LYS[B] |
| | GLU[A] | GLU[A] | ASP[A] | HIS |
| | GLY | ASP[A] | LYS[B] | ARG[B] |
| | ARG[B] | ARG[B] | ARG[B] | ASN |
| | ASP[A] | GLN | GLN | GLN |
| | PRO | ASN | PRO | ASP[A] |
| | **GLN** | SER | ASN | SER |
| | ASN | PRO | ALA | THR |
| | SER | THR | HIS | GLY |
| | THR | GLY | THR | GLU[A] |
| | HIS | ALA | SER | CYS |
| | ALA | HIS | VAL | **ALA** |
| | TYR | TYR | **GLY** | TRP |
| | TRP | TRP | MET | MET |
| | MET | CYS | CYS | VAL |
| | VAL | MET | ILE | TYR |
| | PHE | VAL | LEU | PRO |
| | LEU | ILE | TYR | ILE |
| | ILE | PHE | PHE | LEU |
| Most hydrophobic | CYS | LEU | TRP | PHE |
| Max energy [kcal/mol] | 1.34 | -0.07 | 2.02 | 5.30 |
| Min energy [kcal/mol] | -3.48 | -4.39 | -1.85 | -2.13 |
| Energy range [kcal/mol] | 4.82 | 4.32 | 3.87 | 7.43 |

[A] Acidic AA, [B] Basic AA

Assessment of the HS scales was carried out on small proteins and polypeptide segments as a benchmark for the possibility of the HS to predict the native fold of larger proteins. Seven test proteins similar to ones that are commonly used in state-of-the-art atomistic simulations [5,41] were picked from five different protein classes: transcription, *de novo*, viral, nuclear, and antimicrobial. Their length varies from 10 to 42 AAs, with combinations of three different structural motifs: α-helices, β-sheets, and random coils. PDB entries of the proteins and their properties are given in Table 3. In addition, it should be noted that 1vda includes a combination of a short α-helix and a terminal random coil which is partially stabilized by residue side-chain interactions, thus occasionally forms a hydrophobic core (HC) when the random coil approaches the helical motif. Besides the length, the main difference between the two β-hairpin proteins (1uao and 1u6v) is that 1uao includes acidic residues while 1u6v does not. Both 2bn6 and 2jr8 are mainly helical proteins where 2jr8 forms a long helical segment, while 2bn6 forms two small helical segments that interact via RR contacts and form a HC. 1fme comprises of a combination of α-helical segment and a short β-sheet held together via RR interactions.

**Table 3: Test proteins**

| PDB entry | 1vda | 1uao | 1u6v | 2jof | 2bn6 | 2jr8 | 1fme |
|---|---|---|---|---|---|---|---|
| Protein class | Transcription | *De novo* | Viral | *De novo* | Nuclear | Antimicrobial | *De novo* |
| Length [AA] | 23 | 10 | 17 | 20 | 33 | 42 | 28 |
| Motifs | α-helix random coil | β-hairpin | β-hairpin | α-helix random coil | 2 α-helices | α-helix random coil | α-helix β-sheet |
| Hydrophobic core | Yes/No | No | No | Yes | Yes | No | Yes |

EQUILIBRATION AND FORCE-FIELD TUNING. The force fields were first tested for their ability to preserve the folded structure of the protein in order to assess the stability of the folded state at standard conditions for the length of a standard simulation, and minimally fine-tune the force field in order to maintain the folded state. All equilibration tests begin with structures taken from the PDB. Ten independent MMC trajectories for each HS and protein were used. The PDB reference structures and the resulting trajectories of the equilibration simulations may be found in supplementary material file SM1_equilibration_trajectories.zip. Figure 2 shows the original PDB structure, followed by snapshots of the equilibrated structure using the original HSs (M0, M1, V0, or V1), and after FF tuning, as discussed below. As can be seen, from Figures 2a and 2b, overall, relatively good preservation of protein structure was observed for all HSs, and particularly for M1. Nonetheless, partial unfolding was seen in some cases (i.e., 1u6v, 2jof, 2bn6, and 2ir8), particularly for the vector force fields. Qualitative evaluation of the resulting protein structures during equilibration and after tuning for the different HSs is given in Table 4. Most notably, tuning prevented denaturation for the two vector HSs.

Figure 2b shows that for the original HSs, noticeable deformations were observed mainly for the vector HSs, while the matrix HSs demonstrated relatively good structural stability. While in general all-negative HSs (Figure S1A) demonstrate better packing of the overall structure and an excellent preservation of the protein native structure, the main problem is that deep energy wells give rise to overly stable conformations with any type of RR contact, even the non-native ones. The sticky-termini phenomenon that is observed for all-negative HSs is an excellent example of the problematic energy scale of such HSs, as sticky-termini are practically not observed in PDB. The observations in Figure 2 are in good agreement with the hydrophobic-polar model [42], where contacts with polar residues is not favorable, implying that all-negative HSs might be inaccurate as they presume negative contact energy for polar residues. Specifically, sticky termini were observed for the matrix HSs and were not observed for the vector HSs. Flipped termini (applicable only to 1uao and 1u6v) were observed in all HSs with the exception of M1. Spurious helices and broken HBs (not necessarily of helical segments) were observed for the vector HSs, while broken helices were observed for V1. The formation of a spurious HC was observed in all HSs with the exception of V0. It is important to emphasize that inaccurate FFs are the common reason for structural deformations and for the formation of spurious HCs, as is confirmed by previous reports [43–45]. Detailed evaluation of each HS is provided in the SI.

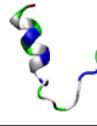

**Figure 2:** Protein folded structures before and after force field tuning. (a) Reference structures taken from the protein data bank (PDB). (b) Equilibrated structures for each hydrophobicity scale (HS) before tuning (M0, M1, V0, V1). (c) Equilibrated structures for each hydrophobicity scale after force tuning. Tuning ensures the same energy range for the different HSs and compensates for energy offsets caused by the different experimental approaches for the determination of the HSs. White segments are hydrophobic, green segments are polar, red segments are acidic, and blue segments are basic.

**Table 4: Visual evaluation of protein structures during equilibration for the four HSs before and after tuning.**

|  | M0 | | M1 | | V0 | | V1 | |
|---|---|---|---|---|---|---|---|---|
|  | Before | After | Before | After | Before | After | Before | After |
| Noticeable deformation |  |  |  |  | X |  | X |  |
| Sticky termini | X | X | X | X |  | X |  | X |
| Flipped termini (1uao, 1u6v) | X | X |  | X | X |  | X | X |
| Spurious helices |  |  |  |  | X |  | X |  |
| Broken helices |  |  |  |  |  |  | X |  |
| Broken hydrogen bonds |  |  |  |  | X |  | X |  |
| Spurious hydrophobic cores | X | X | X | X |  | X | X | X |
| Unstable hydrophobic cores | X | X |  |  | X | X | X | X |
| Denaturation |  |  |  |  | X |  | X |  |

It is evident that the different HSs demonstrate different structural behavior of proteins during equilibration. In some cases, the observed structural deviations might be a result of normal fluctuations [5,41], possibly due to accessible metastable conformations near the native conformation. Since the different scales were determined from different experimental methods, their resulting range of energies and hydrophobicity differ. Tuning of the HSs was attempted to better preserve the structures, but more importantly to determine if there is a set criteria for RR-based HSs that could be used towards the development a universal force field. The HSs were normalized in the following manner: 1) zero hydrophobicity was defined and a corresponding offset was introduced to the HS, and 2) all HSs were scaled uniformly to the same maximal and minimal energy. As seen in Table 2, the zero energy differs between the four HS, being GLN, LYS, GLY, and ALA, respectively. To find a uniform zero we tested three HS cases: 1) all negative, 2) GLY=0, and 3) charged acid/base (A/B) AAs $\geq$ 0. The first case is inspired by the all negative HS M1, the second by previous works where GLY served as a reference [21], and the third case utilizes the logic that acidic (A) and basic (B) AAs prefer interfacing with water and not with HC [42], thus their interaction energies tend to be unfavorable, or non-negative.

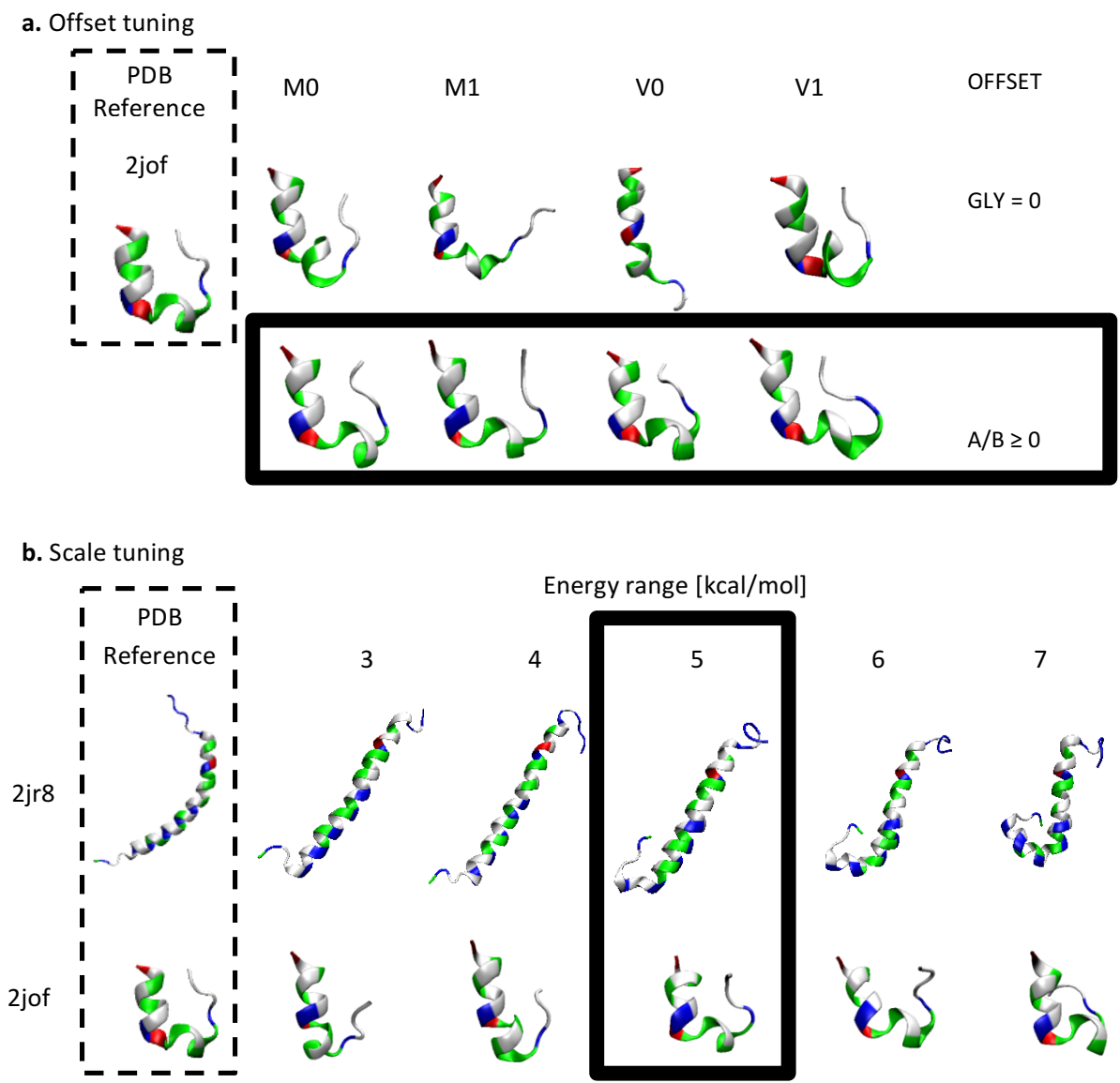

**Figure 3:** Tuning of the hydrophobicity scales via protein equilibration. Finding the optimal offset was done by inspection of structural integrity of proteins under MMC simulation. Representative shapes of protein 2jof for different offsets for all HS (M0,M1,V0,V1) are show in (a). Evidently the optimal offset is where acidic and basic amino acid residues are with non-negative values (A/B≥0). For energy range higher than 5 kcal/mol 2jr8 is deformed, and for energy range less than 5 kcal/mol 2jof loses its structural integrity, leaving 5 kcal/mol as the optimal energy range as shown in (b). Representative structures are shown for M1 HS, with similar tendencies observed for the other HS studied. Additional figures are available in Supporting Information Figures S1A-C, and Figures S2A-D.

Capturing the native fold of a protein requires a fine balance between RR interactions and hydrogen bonding. To determine suitable normalization of the HSs, we set their energy range to a desired test range (from 2 to 8 kcal/mol), by introducing a linear transformation to the HS and preserving the offset constraint for the energy of acidic and basic AAs to be greater than zero in all cases. Because of the large number of scaling experiments (7 ranges x 7 proteins x 4 HS), only four independent trajectories were used for each case (resulting in a total of 784 trajectories). Figure 3a presents selected results for the 2$^{nd}$ and the 3$^{rd}$ offset cases. GLY=0 was used in many HS as the zero reference value [21], however equilibration with such offset lead to significant loss of structure for most of the proteins. The last case where acidic and basic AAs are non-negative was found to give better structural stability than in the 2$^{nd}$ case and better tail fluctuations. Figure 3b shows that the optimal energy range was determined to be near 5 kcal/mol. Higher values encourage undesired bending of the helical protein 2jr8, and lower values fail to preserve shapes of proteins that must maintain a HC, such as 2jof. Interestingly, the optimal range of 5 kcal/mol for the HS scale is very close to the magnitude of HB interaction energy of 4 kcal/mol. The low-entropy structure of the folded protein results from the competition of HBs and local RR interactions, which drives the formation of unique low-entropy structures [46]. HSs with excessively strong (isotropic) RR interactions can destabilize secondary structures that rely on HBs and lead to the formation of non-native hydrophobic or polar cores.

Figure 2c presents selected snapshots of the proteins equilibrated with tuned FFs (compared with the original HS in Figure 2b), and Table 4 provides general observations for the equilibrated structures using the original as well as tuned FFs. Because the change applied to the M0 HS was minor, no practical change of protein structure was observed after tuning. Somewhat better structural fluctuations were observed in all M1 trajectories after the tuning. Sticky termini were still observed, however in fewer cases than before tuning. A noticeable change was observed for both V0 and V1, where the structures were more stable, less deformed, and better preserved. Table 5 presents a summary of energy ranges before and after FF tuning. The row showing weakest A/B ratio gives the offset (and the corresponding AA) applied to the specific HS to maintain the condition A/B≥0, and rows with delta values show the difference in energy between max and min before and after tuning. As explained above, the applied tuning was minor for M0. The tuning was much more significant for M1, V0, and V1 HSs and generally improved the structural stability during equilibration. While some improvements were possible with the tuning procedure, other structural deformations that were observed are the result of intrinsic inaccuracies of the HS for specific proteins.

Table 5: Summary of energy range of the four HSs before and after tuning

|  | M0 | | M1 | | V0 | | V1 | |
| --- | --- | --- | --- | --- | --- | --- | --- | --- |
|  | Original | Tuned | Original | Tuned | Original | Tuned | Original | Tuned |
| max | 1.34 | 1.20 | -0.07 | 0.98 | 2.02 | 1.56 | 5.30 | 2.49 |
| min | -3.48 | -3.80 | -4.39 | -4.02 | -1.85 | -3.44 | -2.13 | -2.51 |
| range | 4.82 | 5.00 | 4.32 | 5.00 | 3.87 | 5.00 | 7.43 | 5.00 |
| Weakest A/B | ASP | ASP | ARG | ARG | ARG | ARG | GLU | GLU |
| Offset [kcal/mol] | 0.18 | 0 | -0.92 | 0 | 0.81 | 0 | 1.60 | 0 |

SIMULATION OF PROTEIN FOLDING. Equilibration of protein structures allowed determining whether the HSs preserved the desired protein structures. Next, simulation of protein folding was performed using the three recipes described in METHODS. Recipe 0 begins with a random conformation, and proceeds according to MMC. The difficulty with recipe 0 is that 250K simulation steps may not be enough to reach equilibrium conformations due to the multitude of metastable conformations. To circumvent this problem, we use Recipes 1 and 2 which accelerate folding simulation in two different ways. It is well known that for a given AA transition, not all ($\varphi,\psi$) pairs on the Ramachandran map are accessible pairs [47–50], thus by sampling conformations from PDB-based histograms in Recipe 1 we eliminate forbidden transitions. Similar to Recipe 0, Recipe 2 also uses standard MMC moves, however, the initial conformation is taken as the most probable according to the same ($\varphi,\psi$) histograms. Recipe 1 folding approach is conceptually similar to that used by Adhikari et al [51].

During the folding simulation, we sample energy and root mean square deviation (RMSD) of C$\alpha$ pairs of protein conformation into 2-dimensional energy-RMSD (ER) histograms. Such histograms allow for a quick determination of the reliability and accuracy of the underlying FF and corresponding HS. As the protein folds, its conformational entropy reduces substantially in favor of the energetically favorable native contacts. Therefore, it is legitimate to assume that conformations with low RMSD should be at energy minima (and indeed has been shown to be the case for two globular proteins [52]), we may easily examine whether this assumption is met by visual inspection of ER distributions and consequently draw conclusions about the accuracy of the corresponding FF and HS. A common method to assess the energy function of proteins is using Z-score, which measures the difference between energies of misfolded protein and that of its native structure in units of the energy standard deviation [53–55]. The main disadvantage of using Z-score is that only the energy values are presented without the corresponding conformational distance (i.e. RMSD). Assessment of energy function via energy and RMSD pairs is much more informative than Z-score and explicitly addresses locations of RMSD minimum and energy minimum. Furthermore, as we will show below, neither criterion alone (low energy or low RMSD) is sufficient to define the native conformation for inaccurate HSs, as non-native conformations may be obtained with similarly low energy or low RMSD.

To quantitatively analyze the ER distributions and evaluate the ability of the HS to predict protein folding, we developed the ER-Score measure. If 50% of the samples with lowest energy and 50% of the samples with lowest RMSD are filtered, the correlation between the two filtered images is defined as the ER-Score:

$$ER = Corr(E, R) = \frac{Cov(E,R)}{\sigma_E \sigma_R} \quad (7)$$

Thus, similarity between the reduced energy and RMSD distributions will give ER-Score near unity, meaning that the HS of interest is accurate. ER-Score values may range from −1 (anti-correlated) to 1 (perfectly correlated). No particular relation is presumed between the energy and RMSD distributions since the score considers significant overlap between the two. Figure 4 (top left) presents 2 conceptual examples of accurate (high ER-score) and inaccurate (low ER-score) distributions. Accurate FFs ensure that energy minima and RMSD minima are found within the same region of the ER distribution [56] while inaccurate FFs allow the energy minima and RMSD minima to be found in different regions of the ER distribution. The advantage of the ER-score is its ability to represent any given ER distribution with a single scalar score, allowing direct comparison between the different ER distributions. Since Z-score is proportional to the conformational energy, we may observe indirectly the distribution of Z-scores of every given case in Figure 4 (by projecting the 2D distribution on the energy axis). Careful examination of the histograms presented in Figure 4 shows that for inaccurate FFs, Z-score is difficult to define because of the spread of low-energy structures. Therefore, our analysis focuses primarily on RMSD minima.

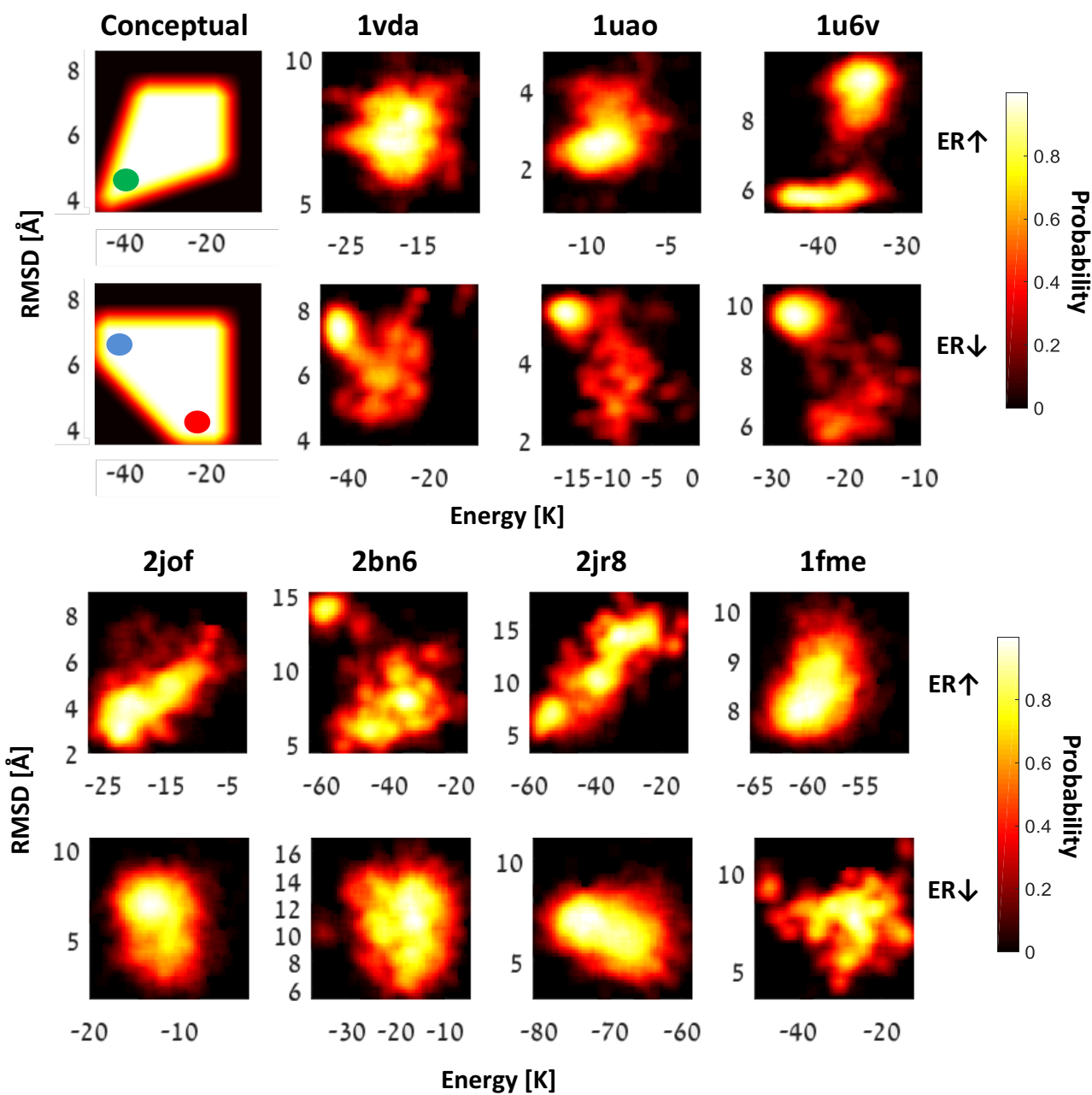

**Figure 4:** Energy-RMSD (ER) distributions of high ER-scores (odd rows) and low ER-scores (even rows). Two conceptual ER distributions are provided to demonstrate the difference between ER distributions with high and low ER-scores: on ER distributions with high ER-score structures with minimum energy and structures with minimum RMSD are found in the same region (green circle) while on ER distributions with low ER-score structures with RMSD minima (red circle), and energy minima (blue circle) reside on different regions of the ER histogram. ER distributions of highest and lowest ER-scores for the seven proteins used in this study are shown. Bright shades represent conformations with higher probability and dark shades represent conformations with lower probability. ER-score is calculated as the correlation between the reduced image (E) with 50% low energy conformations, and the reduced image (R) with 50% low RMSD conformations. Distributions with high ER-score indicate proper force fields, while distributions with low ER-score indicate improper force fields.

ER histograms were sampled from 100 independent trajectories for each test case. ER distributions of all of the cases studied may be found in Figures S3A-D within Supplementary Information. As might be expected, we find that Recipe 0 demonstrated strong dispersions of the ER distributions suggesting that proteins did not converge in most of the cases to their energy minimum in the given simulation timeframe. Recipe 1 trajectories demonstrate narrower ER distribution due to the accelerated nature of their convergence, while Recipe 2 demonstrates small and stable ER islands (e.g., 2bn6 RCP2) in cases that reached energy minima. In cases of weak or missing energy barriers, large and dispersed islands were observed (e.g., 2jof RCP2). Rotated "L"-shaped distributions (e.g., 2bn6 RCP1 V0) were observed for different proteins, and suggest that the energy minimum and the RMSD minimum that were reached belong to different conformations, stressing the risk of using only one of these measures as convergence criterion. In a few cases (2bn6 of M0 and V1) tuning slightly improved ER distributions, however in most of the other cases, tuning demonstrated no practical improvements for the folding experiments, in contrast to the equilibration simulations.

Focusing on stable conformations, we sample ER conformations of only the last 30K steps. Resulting ER distributions of highest and lowest ER-scores are shown in Figure 4 (ER distribution of last 30K steps of every folding simulation may be found in Figures S4A-D). As may be clearly observed, the distributions with high ER-scores are those for which conformations with minimum energy and conformations with minimum RMSD reside in the same region of the ER histogram, with a funnel leading to the low RMSD-energy region. On the other hand, shapes with low ER-scores are those in which conformations with minimum energy and conformations with minimum RMSD reside in different regions of the ER histogram. We can also observe the tendency of the HSs to get trapped in a meta-stable state as may be clearly seen on Figure 4 (row 1 protein 1u6v). In this ER distribution, we observe two low-energy peaks, where one is the native fold (low RMSD) and the other is some metastable fold (high RMSD). Though further analysis is needed, the ER distributions can potentially be used to provide a measure of the entropy, or conformational fluctuations, for both native and metastable folds. Incorrect folds presumably have higher entropy, and thus greater variance around the mean energy and RMSD [52]. Such information can be used in turn towards the development of entropically-driven folding algorithms. Table 6 summarizes information for the ER histograms presented in Figure 4 and provides the overall ER score in each case.

**Table 6: Highest and lowest ER-score experiments**

|  | Protein | 1vda | 1uao | 1u6v | 2jof | 2bn6 | 2jr8 | 1fme |
|---|---|---|---|---|---|---|---|---|
| Highest ER | HS | V0 | V0 | M1 | M1 | V1 | V1 | M1 |
|  | Tuned | Yes | Yes | No | Yes | Yes | No | Yes |
|  | Recipe | 0 | 2 | 2 | 1 | 1 | 1 | 2 |
|  | ER-score | 0.49 | 0.64 | 0.79 | 0.76 | 0.49 | 0.85 | 0.64 |
| Lowest ER | HS | M1 | M1 | V0 | V0 | V0 | V1 | V1 |
|  | Tuned | Yes | No | Yes | No | No | Yes | Yes |
|  | Recipe | 1 | 1 | 1 | 2 | 0 | 2 | 1 |
|  | ER-score | 0 | -0.08 | 0.05 | 0.2 | 0.26 | 0.22 | 0.25 |

So far, we focused on ER-scores for specific proteins; however, it is interesting to observe whether there are overall ER-score trends of the given HS before and after tuning. Table 7 presents mean ER-scores calculated for the last 30K steps of folding simulations before and after tuning for each of the HSs. The scores were calculated as averages of all the sub-scores for the seven proteins used in this study, weighted by the length of every protein (in AA units). ER-scores for the specific proteins may be found Table S1 within Supplementary Information. On average, tuning did not improve folding performance, and in some cases even ended up with worse ER-scores. An interesting conclusion may be drawn here: improvement of equilibration performance does not necessarily yield improved folding performance. Thus, testing HS according to their capacity to maintain a given conformation does not guarantee a successful HS for folding simulation.

Despite the relatively low mean ER-scores, the HSs demonstrate very high ER-scores for specific proteins. The clear tendency of a specific HS to correctly fold specific proteins suggests that a unique and universal RR-based HS may not exist in a *constant* form, and that the hypothetical universal HS must have a dynamic and environment dependent nature, i.e. the contact-energies given within the HS matrix are variable and depend on the surrounding environment. Evidently, these results are in excellent agreement with an early report by Zhang and Kim [57], who have developed a HS that is a function of the specific conformation of the residues of interest. However, their HS depends on the final protein conformation, which makes it difficult to use in prediction-based simulations where the final protein conformation is unknown.

**Table 7: Mean ER-Scores before and after tuning**

|      | M0       |       | M1       |       | V0       |       | V1       |       |
|------|----------|-------|----------|-------|----------|-------|----------|-------|
|      | Original | Tuned | Original | Tuned | Original | Tuned | Original | Tuned |
| RCP0 | 0.46     | 0.35  | 0.45     | 0.47  | 0.37     | 0.43  | 0.41     | 0.41  |
| RCP1 | 0.40     | 0.42  | 0.38     | 0.37  | 0.39     | 0.37  | 0.50     | 0.40  |
| RCP2 | 0.42     | 0.46  | 0.47     | 0.47  | 0.34     | 0.40  | 0.39     | 0.33  |

In addition to the energy-based Z-score, RMSD-based scoring is conventionally used to evaluate folding convergence [58]. Therefore, we examine whether HS that successfully converge to structures with lowest RMSD are necessarily those with highest ER-scores. Figure 5 presents results of successful protein folding trajectories with the specific HS that were filtered according to lowest RMSD during the last 30K steps with the corresponding HS, recipe, and trajectory. Comparison of HS with highest ER-score (Table 6) with HS with lowest RMSD (Figure 5) shows that these HSs are not necessarily the same, suggesting that the HS that reach lowest RMSD converge to meta-stable structures with relatively high surrounding energy barriers and is another side-effect of inaccurate HS.

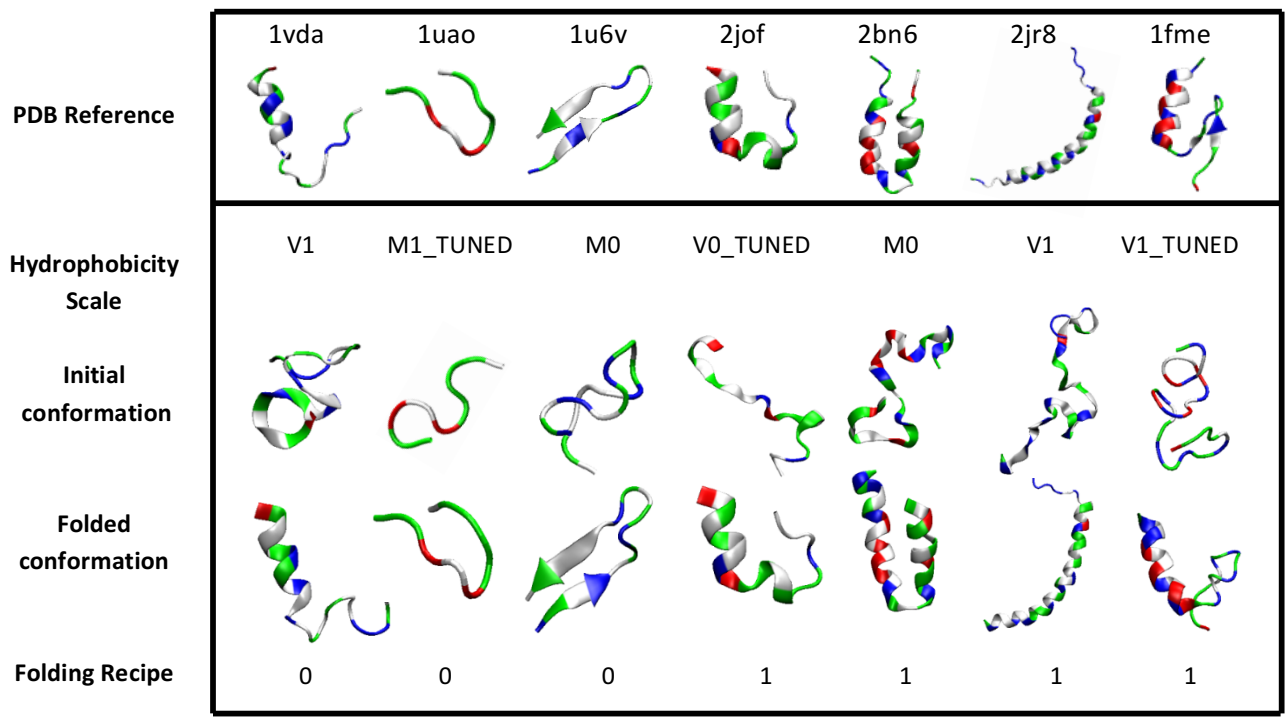

**Figure 5:** Examples of *in silico* folding of protein structures from random initial structures. Detailed animations/trajectories in PDB format may be found in Supplementary Material file SM2_successful_trajectories.zip.

**CONCLUSIONS**

Four hydrophobicity scales (HSs) were studied via equilibration and folding simulations. The HSs were transformed to residue-residue (RR) contact energy matrices and consequently were used as part of the simulation force field (FF). Equilibration MMC simulations showed that some of the HSs in their original form did not successfully preserve protein structures, hence requiring tuning of the HS. For the tuning, we first determined the optimal offset and then the optimal energy range of the HS. Optimal offset was achieved when the hydrophobicity value of weakest acidic/basic amino acid was set to zero. An energy range of 5 kcal/mol, close to the magnitude of single hydrogen-bond on the polypeptide backbone, was found to be optimal for the energy ranges, indicating competition between residue-residue interactions and hydrogen bonding.

An accurate FF should ensure that structures of minimum energy and structures with minimum RMSD are found in the same region of the Energy-RMSD (ER) histogram while inaccurate FFs allow structures with RMSD minima and energy minima to reside on different regions of the ER histogram. For the analysis of ER distributions, the ER-Score was defined as the correlation between conformations with minimum energy and conformations with minimum RMSD. Average scores ranging from 0.33 to 0.5 were observed for the HSs used in this study. Folding experiments showed that the tuning procedure did not always improved ER-Scores, and in some cases even made it worse, pointing to the dissonance between a HS developed for the purpose of maintaining a folded structure versus folding. Although less informative than the distribution, the ER-score provides a single scalar, allowing direct comparison between the different ER distributions. It was shown that the different HSs that successfully fold some proteins from random initial conditions face difficulties folding other proteins (which will only be compounded for larger proteins), reinforcing the conviction that a unique and universal HS may not exist in a constant form. Therefore, a hypothetical universal HS must carry a dynamic and environment dependent nature, i.e. variable contact energies of the HS matrix. This might explain the existence of the many determined HSs.

Simulations in this study were carried out on a consumer-line PC with Intel® Core™ i5-6400 (4 cores) CPU and 8GB RAM memory. Equilibration simulations (10 trajectories, 100K moves/trajectory) lasted approximately a few minutes up to 1 hour for single protein on a single CPU core. Folding simulations (100 trajectories, 350K moves/trajectory) lasted approximately from 0.5 to 35 hours for a single protein on a single CPU core. The remarkable performance is a result of the optimal BRAD representation together with URF coordinates used in this study (described in METHODS section). BRAD together with URF allowed the simulation of proteins in reasonable time using computer configuration that is available practically everywhere and without the need for supercomputers. This efficient computational method potentially open the doors for new studies that aim to develop single and accurate HSs for protein structure prediction.

## ACKNOWLEDGMENT

This work was funded in part by the Israel Science Foundation Grant No. 265/16.

## REFERENCES

(1) Dill, K. A.; MacCallum, J. L. The Protein-Folding Problem, 50 Years on. *Science* **2012**, *338*, 1042–1046.
(2) Dorn, M.; e Silva, M. B.; Buriol, L. S.; Lamb, L. C. Three-Dimensional Protein Structure Prediction: Methods and Computational Strategies. *Comput. Biol. Chem.* **2014**, *53*, 251–276.

(3) Kim, D. E.; Chivian, D.; Baker, D. Protein Structure Prediction and Analysis Using the Robetta Server. *Nucleic Acids Res.* **2004**, *32*, W526–W531.
(4) Berman, H. M.; Westbrook, J.; Feng, Z.; Gilliland, G.; Bhat, T. N.; Weissig, H.; Shindyalov, I. N.; Bourne, P. E. Www.rcsb.org The Protein Data Bank. *Nucleic Acids Res.* **2000**, *28*, 235–242.
(5) Lindorff-Larsen, K.; Piana, S.; Dror, R. O.; Shaw, D. E. How Fast-Folding Proteins Fold. *Science* **2011**, *334*, 517–520.
(6) Brylinski, M. Is the Growth Rate of Protein Data Bank Sufficient to Solve the Protein Structure Prediction Problem Using Template-Based Modeling? *Bio-Algorithms Med-Syst.* **2015**, *11*, 1–7.
(7) Dill, K. A.; Privalov, P. L.; Gill, S. J.; Murphy, K. P. The Meaning of Hydrophobicity. *Science* **1990**, *250* (4978), 297–299.
(8) Young, T. An Essay on the Cohesion of Fluids. *Philos. Trans. R. Soc. Lond.* **1805**, *95*, 65–87.
(9) Wenzel, R. N. Resistance of Solid Surfaces to Wetting by Water. *Ind. Eng. Chem.* **1936**, *28*, 988–994.
(10) A. B. D. Cassie; S. Baxter. Wettability of Porous Surfaces. *Trans. Faraday Soc.* **1944**.
(11) Pauling, L.; Corey, R. B.; Branson, H. R. The Structure of Proteins: Two Hydrogen-Bonded Helical Configurations of the Polypeptide Chain. *Proc. Natl. Acad. Sci.* **1951**, *37*, 205–211.
(12) Zimmerman, J.; Eliezer, N.; Simha, R. The Characterization of Amino Acid Sequences in Proteins by Statistical Methods. *J. Theor. Biol.* **1968**, *21* (2), 170–201.
(13) Aboderin, A. A. An Empirical Hydrophobicity Scale for α-Amino-Acids and Some of Its Applications. *Int. J. Biochem.* **1971**, *2* (11), 537–544.
(14) Bull, H. B.; Breese, K. Surface Tension of Amino Acid Solutions: A Hydrophobicity Scale of the Amino Acid Residues. *Arch. Biochem. Biophys.* **1974**, *161* (2), 665–670.
(15) Tanford, C. The Hydrophobic Effect and the Organization of Living Matter. *Science* **1978**, *200* (4345), 1012–1018.
(16) Alves, N. A.; Aleksenko, V.; Hansmann, U. H. A Simple Hydrophobicity-Based Score for Profiling Protein Structures. *J. Phys. Condens. Matter* **2005**, *17* (18), S1595.
(17) Cozzini, P.; Spyrakis, F. Hydrophobicity in Drug Design. *Int. Union Pure Appl. Chem.* **2006**.
(18) Chowriappa, P.; Dua, S.; Kanno, J.; Thompson, H. W. Protein Structure Classification Based on Conserved Hydrophobic Residues. *IEEE/ACM Trans. Comput. Biol. Bioinform.* **2009**, *6* (4), 639–651.
(19) Giacometti, J.; Josić, D. Protein and Peptide Separations. In *Liquid Chromatography, 1st Edition, Applications/Fanali, Salvatore*; 2013; pp 149–184.
(20) Mageswari, R.; Srinivasa Rao, K.; Sivakumar, K. Prediction of Hydrophobic Core Using Contact Map and Minimal Connected Dominating Set. *Indian Journal of Science* **2015**, *13* (37), 24–28.
(21) Cornette, J. L.; Cease, K. B.; Margalit, H.; Spouge, J. L.; Berzofsky, J. A.; DeLisi, C. Hydrophobicity Scales and Computational Techniques for Detecting Amphipathic Structures in Proteins. *J. Mol. Biol.* **1987**, *195* (3), 659–685.
(22) Simm, S.; Einloft, J.; Mirus, O.; Schleiff, E. 50 Years of Amino Acid Hydrophobicity Scales: Revisiting the Capacity for Peptide Classification. *Biol. Res.* **2016**, *49* (1), 31.
(23) Punta, M.; Maritan, A. A Knowledge-Based Scale for Amino Acid Membrane Propensity. *Proteins Struct. Funct. Bioinforma.* **2003**, *50* (1), 114–121.
(24) Madeira, P. P.; Bessa, A.; Álvares-Ribeiro, L.; Raquel Aires-Barros, M.; Rodrigues, A. E.; Uversky, V. N.; Zaslavsky, B. Y. Amino Acid/water Interactions Study: A New Amino Acid Scale. *J. Biomol. Struct. Dyn.* **2014**, *32* (6), 959–968.
(25) Chothia, C. The Nature of the Accessible and Buried Surfaces in Proteins. *J. Mol. Biol.* **1976**, *105* (1), 1–12.
(26) Engelman, D.; Steitz, T.; Goldman, A. Identifying Nonpolar Transbilayer Helices in Amino Acid Sequences of Membrane Proteins. *Annu. Rev. Biophys. Biophys. Chem.* **1986**, *15* (1), 321–353.

(27) Kapcha, L. H.; Rossky, P. J. A Simple Atomic-Level Hydrophobicity Scale Reveals Protein Interfacial Structure. *J. Mol. Biol.* **2014**, *426* (2), 484–498.
(28) Miyazawa, S.; Jernigan, R. L. Residue–residue Potentials with a Favorable Contact Pair Term and an Unfavorable High Packing Density Term, for Simulation and Threading. *J. Mol. Biol.* **1996**, *256* (3), 623–644.
(29) Berrera, M.; Molinari, H.; Fogolari, F. Amino Acid Empirical Contact Energy Definitions for Fold Recognition in the Space of Contact Maps. *BMC Bioinformatics* **2003**, *4* (1), 8.
(30) Wimley, W. C.; White, S. H. Experimentally Determined Hydrophobicity Scale for Proteins at Membrane Interfaces. *Nat. Struct. Mol. Biol.* **1996**, *3* (10), 842–848.
(31) Moon, C. P.; Fleming, K. G. Side-Chain Hydrophobicity Scale Derived from Transmembrane Protein Folding into Lipid Bilayers. *Proc. Natl. Acad. Sci.* **2011**, *108* (25), 10174–10177.
(32) Metropolis, N.; Ulam, S. The Monte Carlo Method. *J. Am. Stat. Assoc.* **1949**, *44*, 335–341.
(33) Metropolis, N.; Rosenbluth, A. W.; Rosenbluth, M. N.; Teller, A. H.; Teller, E. Equation of State Calculations by Fast Computing Machines. *J. Chem. Phys.* **1953**, *21*, 1087–1092.
(34) Humphrey, W.; Dalke, A.; Schulten, K. VMD: Visual Molecular Dynamics www.ks.uiuc.edu/Research/vmd. *J. Mol. Graph.* **1996**, *14*, 33–38.
(35) Ramachandran, G. N.; Ramakrishnan, C.; Sasisekharan, V. Stereochemistry of Polypeptide Chain Configurations. *J. Mol. Biol.* **1963**, *7*, 95–99.
(36) Ramachandran, G. N. Conformation of Polypeptides and Proteins. *Adv. Protein Chem.* **1968**, *23*, 283.
(37) Flory, P. J. Spatial Configuration of Macromolecular Chains. *Br. Polym. J.* **1976**, *8*, 1–10.
(38) Jorgensen, W. L.; Tirado-Rives, J. The OPLS [optimized Potentials for Liquid Simulations] Potential Functions for Proteins, Energy Minimizations for Crystals of Cyclic Peptides and Crambin. *J. Am. Chem. Soc.* **1988**, *110* (6), 1657–1666.
(39) Jorgensen, W. L.; Maxwell, D. S.; Tirado-Rives, J. Development and Testing of the OPLS All-Atom Force Field on Conformational Energetics and Properties of Organic Liquids. *J. Am. Chem. Soc.* **1996**, *118* (45), 11225–11236.
(40) Haimov, B.; Srebnik, S. A Closer Look into the α-Helix Basin. *Sci. Rep.* **2016**, *6*, 38341.
(41) Adhikari, A. N.; Freed, K. F.; Sosnick, T. R. Simplified Protein Models Can Rival All Atom Simulations in Predicting Folding Pathways and Structure. *Phys. Rev. Lett.* **2013**, *111*, 028103.
(42) Thomas, P. D.; Dill, K. A. Statistical Potentials Extracted from Protein Structures: How Accurate Are They? *J. Mol. Biol.* **1996**, *257* (2), 457–469.
(43) Beauchamp, K. A.; Lin, Y.-S.; Das, R.; Pande, V. S. Are Protein Force Fields Getting Better? A Systematic Benchmark on 524 Diverse NMR Measurements. *J. Chem. Theory Comput.* **2012**, *8* (4), 1409–1414.
(44) Lindorff-Larsen, K.; Maragakis, P.; Piana, S.; Eastwood, M. P.; Dror, R. O.; Shaw, D. E. Systematic Validation of Protein Force Fields against Experimental Data. *PloS One* **2012**, *7* (2), e32131.
(45) Piana, S.; Klepeis, J. L.; Shaw, D. E. Assessing the Accuracy of Physical Models Used in Protein-Folding Simulations: Quantitative Evidence from Long Molecular Dynamics Simulations. *Curr. Opin. Struct. Biol.* **2014**, *24*, 98–105.
(46) Srebnik, S.; Chakraborty, A. K.; Shakhnovich, E. I. Adsorption-Freezing Transition for Random Heteropolymers near Disordered 2d Manifolds due to "pattern Matching." *Phys. Rev. Lett.* **1996**, *77* (15), 3157.
(47) Jha, A. K.; Colubri, A.; Zaman, M. H.; Koide, S.; Sosnick, T. R.; Freed, K. F. Helix, Sheet, and Polyproline II Frequencies and Strong Nearest Neighbor Effects in a Restricted Coil Library. *Biochemistry (Mosc.)* **2005**, *44*, 9691–9702.

(48) Ting, D.; Wang, G.; Shapovalov, M.; Mitra, R.; Jordan, M. I.; Dunbrack Jr, R. L. Neighbor-Dependent Ramachandran Probability Distributions of Amino Acids Developed from a Hierarchical Dirichlet Process Model. *PLoS Comput Biol* **2010**, *6*, e1000763.

(49) Carugo, O.; Djinovic-Carugo, K. Half a Century of Ramachandran Plots. *Acta Crystallogr. D Biol. Crystallogr.* **2013**, *69*, 1333–1341.

(50) Carrascoza, F.; Zaric, S.; Silaghi-Dumitrescu, R. Computational Study of Protein Secondary Structure Elements: Ramachandran Plots Revisited. *J. Mol. Graph. Model.* **2014**, *50*, 125–133.

(51) Adhikari, A. N.; Freed, K. F.; Sosnick, T. R. Simplified Protein Models: Predicting Folding Pathways and Structure Using Amino Acid Sequences. *Phys. Rev. Lett.* **2013**, *111*, 028103.

(52) Bowman, G. R.; Pande, V. S. The Roles of Entropy and Kinetics in Structure Prediction. *PloS One* **2009**, *4* (6), e5840.

(53) Sippl, M. J. Knowledge-Based Potentials for Proteins. *Curr. Opin. Struct. Biol.* **1995**, *5* (2), 229–235.

(54) Zhang, L.; Skolnick, J. What Should the Z-Score of Native Protein Structures Be? *Protein Sci.* **1998**, *7* (5), 1201–1207.

(55) Vendruscolo, M. Assessment of the Quality of Energy Functions for Protein Folding by Using a Criterion Derived with the Help of the Noisy Go Model. *J. Biol. Phys.* **2001**, *27* (2), 205–215.

(56) MacDonald, J. T.; Kelley, L. A.; Freemont, P. S. Validating a Coarse-Grained Potential Energy Function through Protein Loop Modelling. *PloS One* **2013**, *8* (6), e65770.

(57) Zhang, C.; Kim, S.-H. Environment-Dependent Residue Contact Energies for Proteins. *Proc. Natl. Acad. Sci.* **2000**, *97* (6), 2550–2555.

(58) Kufareva, I.; Abagyan, R. Methods of Protein Structure Comparison. *Homol. Model. Methods Protoc.* **2012**, 231–257.